\documentclass[twocolumn]{aa}
\usepackage{graphicx}
\usepackage{graphics}
\usepackage{txfonts}
\usepackage{epsfig}
\usepackage{natbib}
\usepackage{times}
\bibpunct{(}{)}{;}{a}{}{,}

\def \INT {\emph{INTEGRAL}}
\def \ISGRI {\emph{ISGRI}}
\def \IGR {IGR\,J16393$-$4643}
\def \XMM {\emph{XMM-Newton}}
\def \EPIC {\emph{EPIC}}
\def \PN {\emph{EPIC/PN}}    
\def \MOS {\emph{EPIC/MOS}}
\def \MOSone {\emph{EPIC/MOS1}}
\def \MOStwo {\emph{EPIC/MOS2}}
\def \ASCA {\emph{ASCA}}

\begin{document}

\title{IGR\,J16393$-$4643: a new heavily-obscured X-ray pulsar}

\author{A.~Bodaghee \inst{1,2}, 
R.~Walter \inst{1,2}, 
J.A.~Zurita~Heras \inst{1,2}, 
A.J.~Bird \inst{3}, 
T.J.-L.~Courvoisier \inst{1,2},
A.~Malizia \inst{4}, 
R.~Terrier \inst{5,6}, 
P.~Ubertini \inst{7}
}

\authorrunning{Bodaghee~A.~et~al.}
	\offprints{arash.bodaghee@obs.unige.ch}

\institute{
	\INT\ Science Data Centre, Chemin d'Ecogia 16, CH--1290 Versoix, Switzerland
    \and
	Observatoire de Gen\`eve, Chemin des Maillettes 51, CH--1290 Sauverny, Switzerland
	\and
	School of Physics and Astronomy, University of Southampton, Highfield, Southampton, SO17 1BJ, UK
	\and
	IASF/CNR, Via Gobetti 101, I--40129 Bologna, Italy
	\and
	CEA-Saclay, DAPNIA/Service d'Astrophysique, F--91191 Gif sur Yvette Cedex, France
	\and
	F\'ed\'eration de recherche APC, Coll\`ege de France 11, place Marcelin Berthelot, F--75231 Paris, France
	\and
	IASF/CNR, Via Fosso del Cavaliere 100, I--00133 Rome, Italy
	}
	\date{Received / Accepted }

\abstract{An analysis of the high-energy emission from \IGR\ ($=$AX\,J1639.0$-$4642) is presented using data from \INT\ and \emph{XMM-Newton}. The
source is persistent in the 20--40 keV band at an average flux of $5.1\times10^{-11}$ergs\,cm$^{-2}$\,s$^{-1}$, with variations
in intensity by at least an order of magnitude. A pulse period of 912.0$\pm$0.1 s was discovered in the \ISGRI\ and \EPIC\ light curves.
The source spectrum is a strongly-absorbed ($N_{\mathrm{H}}=(2.5\pm0.2)\times10^{23}$\,cm$^{-2}$) power law that features a high-energy cutoff 
above 10\,keV. Two iron emission lines at 6.4 and 7.1 keV, an iron absorption edge $\gtrsim$7.1 keV, and a soft excess 
emission of $7\times10^{-15}$ergs\,cm$^{-2}$\,s$^{-1}$ between 0.5--2 keV, are detected in the \EPIC\ spectrum.
The shape of the spectrum does not change with the pulse. Its persistence, pulsation, and spectrum place \IGR\ among
the class of heavily-absorbed HMXBs. The improved position 
from \EPIC\ is R.A. (J2000)$=16^{\mathrm{h}}39^{\mathrm{m}}05.4^{\mathrm{s}}$ and Dec.$=-46^{\circ}42'12''$ ($4''$
uncertainty) which is compatible with that of 2MASS\,J16390535$-$4642137.
}

\maketitle

\section{Introduction}

The \INT\ core program \citep[CP:][]{Win03} routinely devotes observation time to Galactic Plane Scans (GPS) and
Galactic Centre Deep Exposures (GCDE). These numerous snapshots of the Milky Way can be assembled into mosaic images of
long exposure time ($\sim$1 Ms). This gamma-ray view of the galaxy, as collected by
\ISGRI\ \citep{Ube03,Leb03}, enabled \citet{Bir04} to detect 123 sources at a significance above 6$\sigma$. Around 20 of these 
sources are of unknown origin. A good portion of these new sources probably belong to the class of heavily-absorbed 
High Mass X-ray Binaries (HMXBs) that are concentrated along the galactic plane and in the spiral arms.

High-Mass X-ray Binaries are composed of a compact object such as a neutron star 
or a black hole that orbits a massive stellar companion. Depending on the type of companion, known HMXBs can be 
divided into two groups \citep{Par83}. Most HMXBs classified by \citet{Liu00} contain a Be star. These systems are 
usually transient sources with hard spectra. The compact object has a wide orbit which mostly keeps it away from the 
Be star and its disk. Outbursts in these systems are due primarily to the compact object approaching the star and accreting 
matter from the slow, dense stellar wind. The second group of HMXBs features an O or B supergiant star. The orbit of the compact object places it well 
within the stellar wind, so material from the supergiant can be fed directly to the compact object through Bondi 
accretion, or it can pass to the compact object via an accretion disk. The latter mechanism is typically found in 
bright X-ray binaries in which the Roche lobe overflow of gas from the OB star supercedes the flow of accreting matter. 
For less luminous binaries, the OB star does not fill its Roche lobe and the behaviour of the X-ray source is determined 
predominantly by the stellar wind. X-ray emission in supergiant HMXB systems is usually persistent, with flares 
stemming from inhomogeneities in the wind. Neutron stars with strong magnetic fields develop a 
hot spot for accretion which can result in a pulsation. 

\begin{figure*}[!t]  \centering
\psfig{width=6cm,angle=270,file=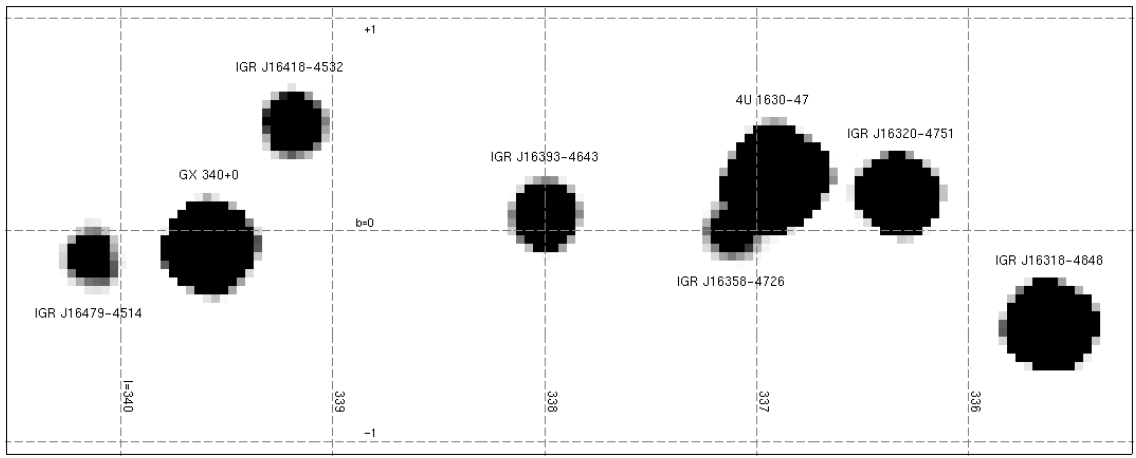}
\caption[]{Intensity map of the Norma Arm in the 20--40 keV band accumulating data from revolutions
30--185 for an effective exposure time of 670 ks. The image is in galactic coordinates and represents an area 2$^{\circ}$ tall by 5$^{\circ}$ wide.
The galactic equator ($b=0$) bisects the image horizontally. Neighboring sources from the catalog of \citet{Bir04} are also shown.}
\label{isgri_mos}
\end{figure*}

The number of persistent and heavily-absorbed HMXBs associated with 
supergiant companions has increased thanks to deep, wide-field observations by \INT\ combined 
with follow-up X-ray monitoring by \XMM\ \citep{Wal03,Rod03,Hil05,Lut05b}. So far, these sources have been detected 
preferentially in the Norma Arm region \citep{Wal04} which features high formation rates of OB supergiant stars. 
It is there that \INT\ uncovered its first new source, IGR\,J16318$-$4848 \citep{Wal03}. The Norma Arm also 
harbors \IGR. This object was initially discovered in the X-ray band and listed as AX\,J163904$-$4642 in the \ASCA\ 
Faint Source Catalog \citep{Sug01}. Its flat power-law slope, its absorption ($N_{\mathrm{H}}=13^{+9}_{-7}\times10^{22}\,$cm$^{-2}$), 
its lack of radio emission, and its position in the galactic plane compelled the authors to classify it as a HMXB. A 
non-thermal radio counterpart was recently detected in the \ASCA\ error box which suggests a dust-enshrouded microquasar 
interpretation for the HMXB \citep{Com04}, and would help explain the 
possible association with the unidentified \emph{EGRET} source 3EG\,J1639$-$4702 \citep{Har99} noted by \citet{Mal04}.
 
We detected \IGR\ in the \ISGRI\ GPS data of the Norma Arm's first visibility period, and we obtained
a follow-up observation with \XMM. The set of \INT\ and \XMM\ data are introduced in
Section\,\ref{obsdata}. In Section\,\ref{count}, the refined X-ray position is used to locate the most
likely counterpart at other wavelengths. Timing and spectral analyses are presented in Sections\,\ref{time} and \ref{spec}, respectively.
Finally, we discuss the nature of the source and we offer our conclusions in Section\,\ref{disc}.

\section{Observations and Data Sets}
\label{obsdata}
\subsection{INTEGRAL Data and Imaging}
\label{intdata}

The \INT\ data consist of all CP pointings during revolutions 30--260, as well as pointings that were public by
January 3, 2005, which had the source within the \ISGRI\ field of view (FOV). To improve the quality of the output mosaic,
pointings with exposure times below 1 ks were ignored. The resulting data set groups roughly 1500 pointings with an
average exposure time of 2 ks each.

Version 4.2 of the \INT\ Offline Scientific Analysis (OSA) software was
used to reduce raw data into images. Source extraction employed version 18 of the \INT\ General Reference Catalog \citep{Ebi03}
selected for objects detected by \ISGRI. These tools are available to the public through
the \INT\ Science Data Centre \citep[ISDC:][]{Cou03}.

Intensity, significance, variance, and exposure mosaic images were constructed from background-subtracted images of individual
pointings. Figure\,\ref{isgri_mos} provides an example of an intensity map of \IGR\
and its vicinity in the 20--40 keV band from the pointings of revolutions
30 to 185. The effective exposure time is 670 ks after correcting for vignetting.
Using this image, we extracted a source location of
R.A. (J2000)$=16^{\mathrm{h}}39^{\mathrm{m}}05^{\mathrm{s}}$ and Dec.$=-46^{\circ}42.3'$ ($26''$ uncertainty) which agrees with and improves the 
\ISGRI\ position of \citet{Bir04}.
The mean flux (20--60 keV) of the source is $0.73\pm0.02$ counts per second (cps), or 4.9 mCrab,
at a significance of 36$\sigma$.
%

\subsection{XMM-Newton Data and Imaging}
\label{xmmdata}

\XMM\ \citep{Jan01,Str01,Tur01} observed \IGR\ on March 21, 2004, from 08:21:15 to 11:41:15 (UT). We used the Science
Analysis System (SAS) v. 6.1.0 software
to analyse the data and to extract the \EPIC\ spectrum. The data were screened for background variability by excluding time
intervals in which the count rate above 10 keV was greater than
a selected threshold (0.33 cps for \MOSone, 0.25 cps for \MOStwo, and 2.4 cps for \PN).
After screening, the effective exposure times were around 8.4, 8, and 7 ks for \MOSone, \MOStwo, and \PN,
respectively.

\begin{figure}[!t]  \centering
\psfig{width=7cm,file=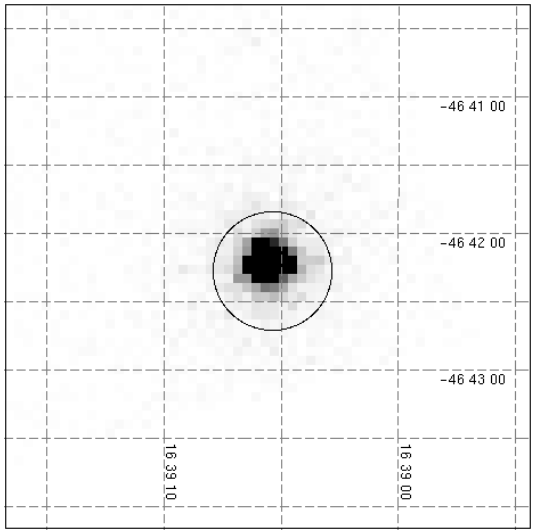}
\caption[]{\MOSone\ image of \IGR\ in the 0.15--12 keV energy range with an effective exposure time of 8.4 ks.
The refined \ISGRI\ error circle from this paper is superimposed.}
\label{xmm_mos}
\end{figure}

\IGR\ is clearly detected in images taken by \EPIC. Figure\,\ref{xmm_mos} presents an image of \IGR\ from the 
\MOSone\ camera. Furthermore, the source
coincides with the updated \ISGRI\ position and error circle from this paper. The refined X-ray
position averaged from \MOS\ and \PN\ is
R.A. (J2000)$=16^{\mathrm{h}}39^{\mathrm{m}}05.4^{\mathrm{s}}$ and Dec.$=-46^{\circ}42'12''$ ($4''$ uncertainty).
%

\section{Counterparts}
\label{count}

We searched for counterparts at other wavelengths and found a single infrared source belonging to the Two 
Microns All-Sky Survey \citep[2MASS:][]{Cut03} in the \XMM\ error box. This potential counterpart, 2MASS\,J16390535$-$4642137, 
is located about $2''$ away from the \XMM\ position at
R.A. (J2000)$=16^{\mathrm{h}}39^{\mathrm{m}}05.36^{\mathrm{s}}$ and Dec.$=-46^{\circ}42'13.7''$ ($0.06''$ uncertainty). 
It appears in the $J$, $H$, and $K_{s}$ bands with magnitudes of
14.63$\pm$0.06, 13.32$\pm$0.04, and 12.78$\pm$0.04, respectively (95\% confidence).

\begin{figure}[!t]  \centering
\psfig{width=6cm,angle=270,file=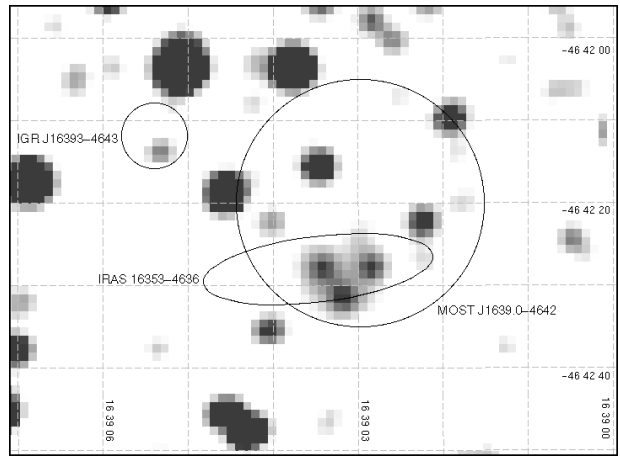}
\caption[]{Image from the 2MASS survey of the region around \IGR\ in the $K_{s}$ band.
The \XMM\ error circle from this paper does not intersect the error boxes of the radio source MOST\,J1639.0$-$4642 and the 
far infrared source IRAS\,16353$-$4636. The only infrared object inside the \XMM\ error circle is 2MASS\,J16390535$-$4642137.}
\label{combi_cp}
\end{figure}

Figure\,\ref{combi_cp} shows that the \XMM\ position and error radius for \IGR\ do not intersect 
the error boxes of the radio source MOST\,J1639.0$-$4642 and the far infrared source IRAS\,16353$-$4636, which are 
the possible counterparts proposed by \citet{Com04}. The tight \XMM\ error circle excludes optical counterparts from 
the USNO-B1 catalog \citep{Mon03}, radio sources from the Vizier database \citep{Och00},
and \emph{ROSAT} sources \citep{Vog99}.

A possible association of \IGR\ with the \emph{EGRET} source 3EG\,J1639$-$4702 \citep{Har99} has been
proposed by \citet{Mal04}. Another \ISGRI\ source, IGR\,J16358$-$4726, and 4U\,1630$-$47 are just outside the 
\emph{EGRET} 95\% error contour. In the $2^{\circ}\times5^{\circ}$ degree section of the Norma Arm presented in Fig.\,\ref{isgri_mos}, 
the probability to observe a counterpart in the 0.56$^{\circ}$ error radius of the \emph{EGRET} source is close to 1, so 
the coincidence with \IGR\ is probably a chance one. 

\section{Timing Analysis}
\label{time}
\subsection{Long-term Variability}
\label{longvar}

Most \ISGRI\ pointings have exposure times that are too brief for a significant detection. Therefore, a mosaic was prepared for
each 3-day spacecraft revolution (from 30--244) in the 20--60 and 60--150 keV bands. We extracted the source flux, error, and 
significance by fitting a Gaussian with a fixed point spread function to the mosaic images. 
Upper limits (3$\sigma$), calculated from the variance maps, are provided when the source is not detected at 
the 4$\sigma$ level in a mosaic image with an effective exposure time above 7 ks. 
There were no detections in mosaic images of the 60--150 keV band so its light curve is omitted.

\begin{figure}[!t]
\psfig{width=7.5cm,angle=0,file=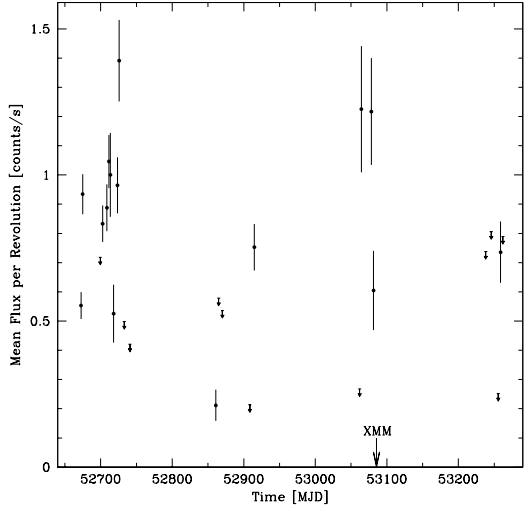} \centering
\caption[]{\ISGRI\ light curve (20--60 keV) of \IGR\ where each point represents the flux averaged over 1 spacecraft 
revolution. Upper limits (3$\sigma$) are provided when the source is not detected for an effective exposure time above 
7 ks. The \XMM\ observation (Fig.\,\ref{xmm_lc}) occurred on MJD 53085.}
\label{isg_lc_rev}
\end{figure}

Figure\,\ref{isg_lc_rev} illustrates the persistence of \IGR\ in the 20--60 keV energy range. Table\,\ref{tab_lc_rev} 
collects the 15 revolutions in which the source is detected in the mosaic image with a significance above 4$\sigma$. 
The average flux in these mosaics is 0.86 cps (5.6 mCrab) with a cumulative exposure time of 1.5 Ms.
The mean flux per revolution varies by a factor of at least 6 from 0.21$\pm$0.05 counts per second (cps) or 
1.4 mCrab in revolution 100 (205 ks), to 1.39$\pm$0.14 cps (9.3 mCrab) in revolution 55 (96 ks). 
When the flux value during revolution 100 is compared to the highest flux registered in a single 2-ks pointing of the 20--60 keV band (4.3$\pm$0.6 cps, 
or 29 mCrab, during MJD 52673.623--52673.641), we find that the source flux varies by a factor larger than 20.

\begin{table}[!b] \centering
        \caption{Source flux value (20--60 keV), significance, average off-axis angle, and exposure time 
	when \IGR\ is detected at the 4$\sigma$ level in the mosaic image of a spacecraft revolution. The median 
	MJD of the visibility period is also given in reference to Fig.\,\ref{isg_lc_rev}.}
        \label{tab_lc_rev}
        \begin{tabular}{cccccc}  \hline
        \noalign{\smallskip}
        \noalign{\smallskip}
        \multicolumn{2}{c}{Visibility Period} & Exp. & Flux  & Sig. & Angle  \\
        \noalign{\smallskip}
          $[$Rev.$]$ & $[$MJD$]$ & $[$ks$]$ & $[$counts$/$s$]$ & $[\sigma]$ & $[^{\circ}]$  \\
        \noalign{\smallskip}
        \noalign{\smallskip}
	\hline
        \noalign{\smallskip}
        \noalign{\smallskip}
	37       & 52672 & 204 & 0.55$\pm$0.05 & 12.1 &  3.8   \\
	38       & 52675 & 92 & 0.93$\pm$0.07 & 13.7 &  3.6   \\
	47       & 52703 & 153 & 0.83$\pm$0.06 & 13.3 &  7.0   \\
	49       & 52709 & 85 & 0.89$\pm$0.08 & 11.1 &  6.2   \\
	50       & 52712 & 104 & 1.05$\pm$0.09 & 11.6 &  9.5   \\
	51       & 52714 & 76 & 1.00$\pm$0.14 & 7.0  &  11.3  \\
	52       & 52718 & 93 & 0.53$\pm$0.10 & 5.3  &  10.1  \\
	54       & 52724 & 100 & 0.96$\pm$0.10 & 10.0  &  9.7   \\
	55       & 52726 & 96 & 1.39$\pm$0.14 & 10.0 &  11.9  \\
	100      & 52861 & 205 & 0.21$\pm$0.05 & 4.0 &  6.9   \\
	118      & 52915 & 146 & 0.75$\pm$0.08 & 9.4  &  10.0  \\ 
	168      & 53064 & 28 & 1.22$\pm$0.22 & 5.7  &  11.5  \\
	173      & 53078 & 27 & 1.22$\pm$0.18 & 6.7  &  9.3   \\ 
	174      & 53081 & 53 & 0.60$\pm$0.14 & 4.5  &  9.3   \\
	233      & 53259 & 72 & 0.74$\pm$0.10 & 7.0  &  9.2   \\
	\noalign{\smallskip}
        \noalign{\smallskip}
        \hline
        \end{tabular} 
\end{table}

\begin{figure}[!t]
\psfig{width=7.5cm,angle=0,file=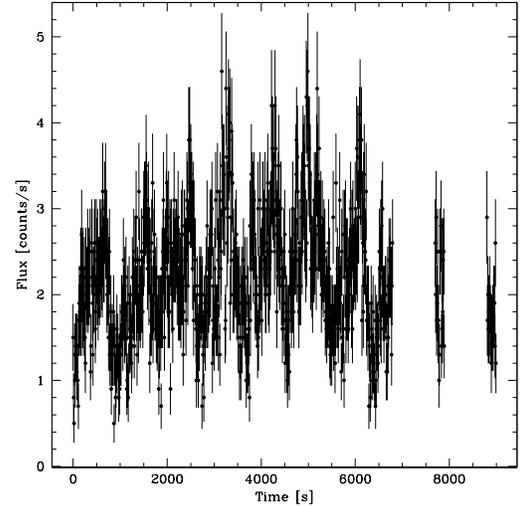} \centering 
\caption[]{\PN\ light curve for \IGR\ in the 2--10 keV energy range with a time resolution of
10 s. The observation starts at MJD 53085.37516.}
\label{xmm_lc}
\end{figure}

\subsection{Pulsations}
\label{shortvar}

\begin{figure}[!t]
\psfig{width=7.5cm,angle=0,file=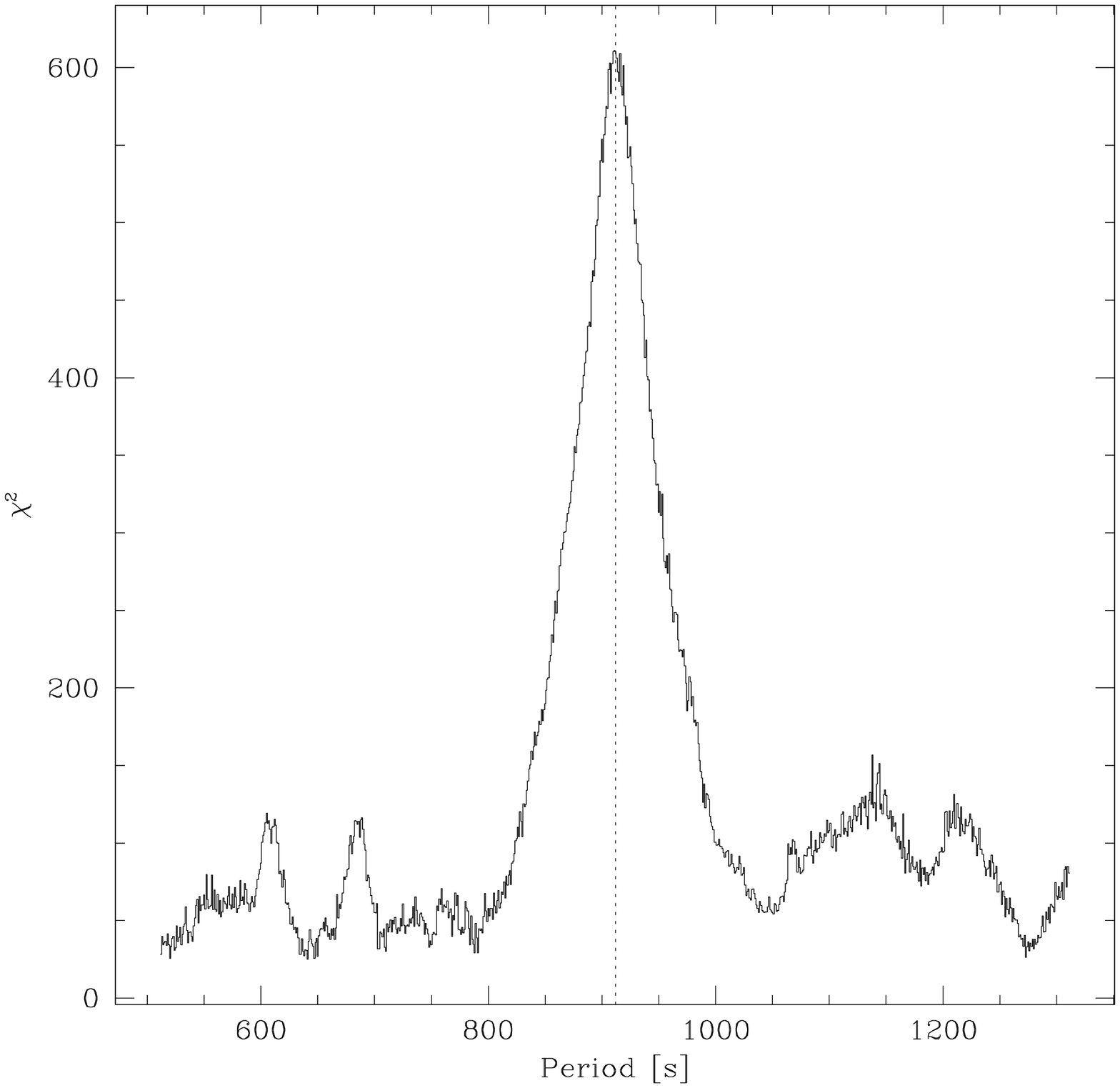} \centering
\psfig{width=7.5cm,angle=0,file=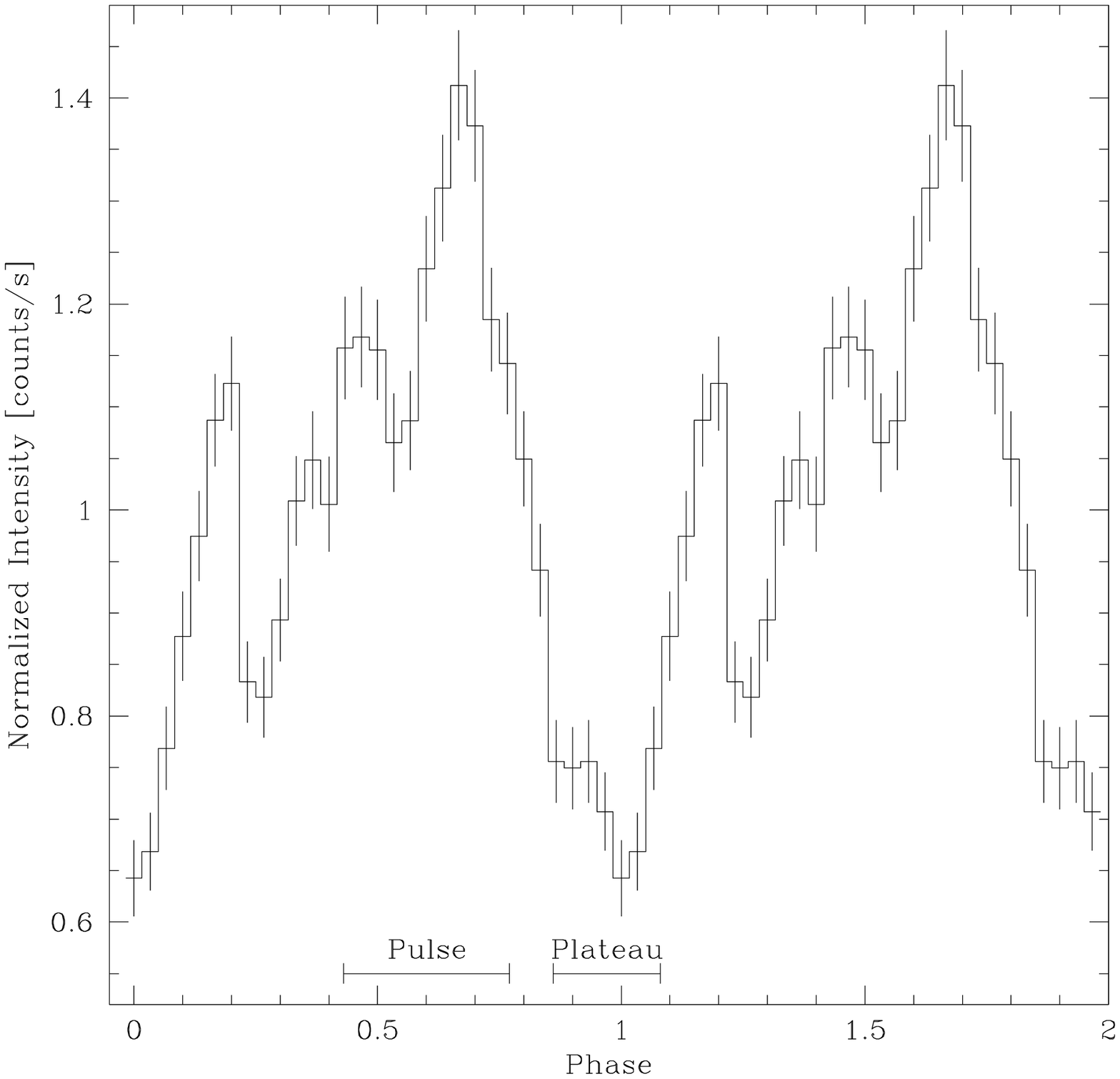}
\caption[]{\emph{Top} (a): Distribution of $\chi^{2}$ for the \PN\ 2--10 keV light curve with 30 bins per period, a resolution of 1 s, 
and centred at 912 s (vertical line). \emph{Bottom} (b): Folded \PN\ light 
curve for a pulse period of 912 s beginning at MJD 53085.38738. The 
intervals indicate the pulse and plateau states of the phase-resolved spectrum.}
\label{xmm_var}
\end{figure}

During the \PN\ observation of 8 ks, the source count rate varied from 1 to 4 cps
(Fig.\,\ref{xmm_lc}). Furthermore, the variations occur periodically.
By searching for modulations in the power spectrum and in the $\chi^{2}$ distribution, we obtain a
period of 912$\pm$5 s at a $\chi^{2}$ of $\sim$600 for 30 bins per period
(Fig.\,\ref{xmm_var}a). The pulse period is determined by the centre of a Gaussian fit to the $\chi^{2}$ distribution, and the 
error is calculated using the procedure developed by \citet{Hor86} on Lomb-Scargle periodograms. The \PN\ pulse profile folded with a 
period of 912 s, illustrated in Fig.\,\ref{xmm_var}b, indicates a pulse fraction of 38$\pm$5\%.

\begin{figure}[!t]
\psfig{width=7.5cm,angle=0,file=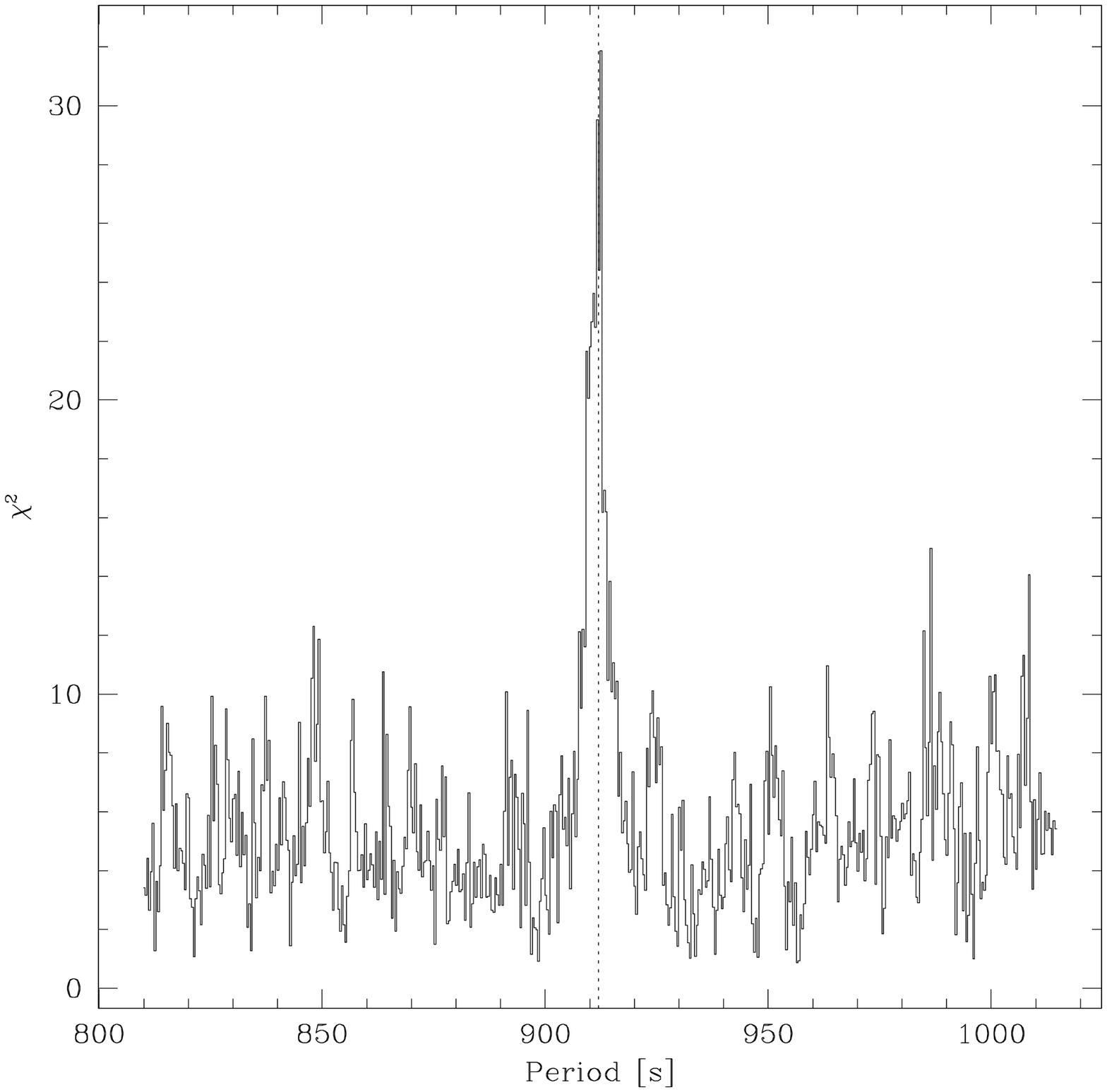} \centering
\psfig{width=7.5cm,angle=0,file=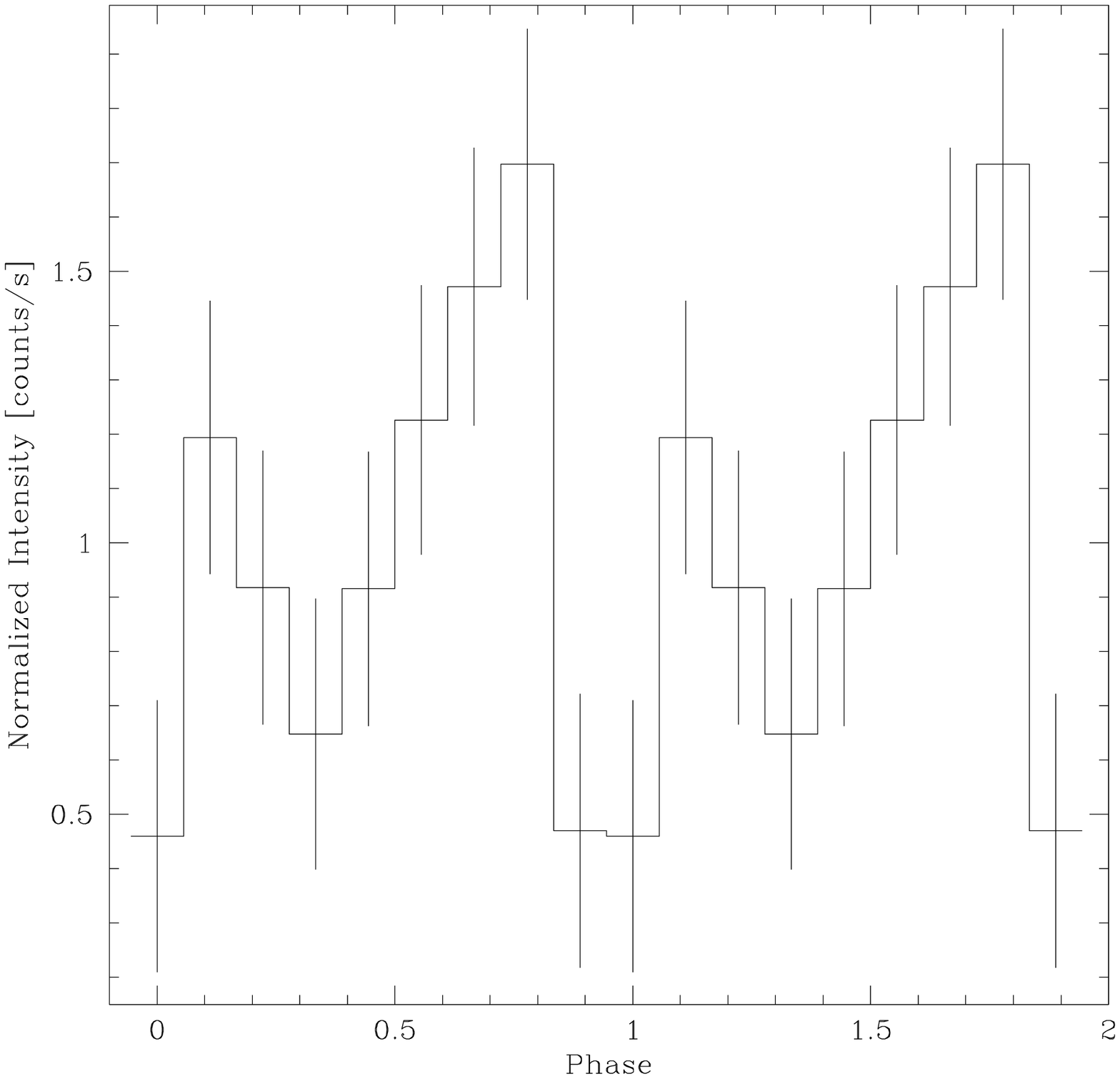}
\caption[]{\emph{Top} (a): Period search ($\chi^{2}$ distribution) on the \ISGRI\ light curve (15--40 keV) of
revolution 38, centred at 912 s (vertical line), with 9 bins per period, and a resolution of 0.4 s. 
\emph{Bottom} (b): Pulse profile of the folded \ISGRI\ light curve (15--40 keV) for a period of 912 s.
The start time is MJD 52674.53833.}
\label{isg_var}
\end{figure}

For \ISGRI, the pulse is best detected in the 100-s light curve of revolution 38 in the 15--40 keV band since
this revolution has enough effective exposure time (50 ks) with low off-axis angles ($\sim4^{\circ}$),
and its light curve is free of gaps that hinder a periodicity search. Figure\,\ref{isg_var}a presents the
$\chi^{2}$ distribution centered at 912 s, with 9 bins per period. The phase diagram is shown in Fig.\,\ref{isg_var}b, and 
the pulse fraction is 57$\pm$24\%. We merged the 100-s light curves (15--30 keV) from revolutions 
37--55, and derived a pulse period and error of 912.0$\pm$0.1 s.

With respect to the pulse fractions, the amplitude does not change significantly as a function of energy. 
Both display a jagged rise to a peak flux followed by a drop. There appear to be primary and secondary peaks in the pulse 
profiles of both instruments at phases of $\sim$0.7 and $\sim$0.2, respectively. Although the \ISGRI\ and \EPIC\ periods are derived from 
observations about 400 days apart, the period is not accurate enough to search for possible variations. 
In addition, the \ISGRI\ light curves of revolutions that are nearly concurrent with the \EPIC\ observation present low source 
significances which makes it difficult to search for a period.

There is an indication in the \EPIC\ light curve (Fig.\,\ref{xmm_lc}) that the source varies
over timescales longer than the pulse period. At the beginning of the observation, the maximum of the pulse is
about 3 cps. This maximum rises to 4 cps about 5 ks after the start time, and then decreases.

We did not detect an orbital period of the order of a few days in the combined \ISGRI\ light curves (15--30 keV, 100-s resolution) 
of detected revolutions between 37 and 55.

\section{Spectral Analysis}
\label{spec}

\subsection{Average Spectrum}

To extract an \ISGRI\ spectrum, we created mosaics accumulating the data from revolutions 37--185 in the
following energy bands; 22.09--25.92, 25.92--30.23, 30.23--40.28, 40.28--51.29, and 51.29--63.26 keV.
These boundaries were chosen to conform with the current response matrices. The source is clearly detected up to
40 keV, so spectral bins at higher energies are ignored. A power law fit to the
\ISGRI\ spectrum returns a relatively steep photon index of 4.5$\pm$0.4 ($\chi_{\nu}^{2}=0.32$).	

Spectral extraction for \PN, \MOSone, and \MOStwo\ relied on single and double events
within a circle of radius 25$''$ centred on the source. To estimate the background, we made an event list from a circle of
equal radius in the same detector and at an equivalent distance from the detector's edge. Channels were configured
to collect 30 counts per bin. An absorbed power law fit to the \EPIC\ spectra has a photon index of 1.0$\pm$0.1 with a 
column density of 3.1$\pm$0.1$\times10^{23}$cm$^{-2}$ ($\chi_{\nu}^{2}=1.15$).

The \EPIC\ spectrum is heavily-absorbed below 4 keV. Iron emission
lines appear at 6.4 keV (Fe\,K$_{\alpha}$) and at 7.1 keV (Fe\,K$_{\beta}$). A discontinuity
in the continuum above 7.1 keV suggests an iron absorption edge. There is an indication of an excess of soft emission between
0.5 and 2 keV. Including a soft blackbody component in an absorbed power law adjusted to the \EPIC\ spectra decreases the 
$\chi^{2}$ by 9 for 680 degrees of freedom (dof), or a 0.5\% probability that this feature is due to chance. Using a partial covering absorption 
instead of a blackbody raises the $\chi^{2}$ by 60 for the same dof (679). A partially-ionised absorber is also insufficient to model the 
soft excess.

Simply fitting an absorbed power law to the spectra from \EPIC\ and \ISGRI\ yields $N_{\mathrm{H}}\sim4\times10^{23}$cm$^{-2}$ and 
$\Gamma\sim2.3$, but the fit is poor ($\chi_{\nu}^{2}=2.1$) with significant residuals in the hard X-rays. 
The \EPIC\ photon index of 1.0$\pm$0.1 is smaller than the one observed
for \ISGRI\ (4.5$\pm$0.4) which indicates a spectral break. Such spectral shapes---a flat power law at low energies with a high-energy cutoff
between 10--20 keV beyond which the slope steepens---are typical of X-ray pulsars \citep{Whi95}.

\begin{figure}[!t]
\psfig{width=5.5cm,angle=270,file=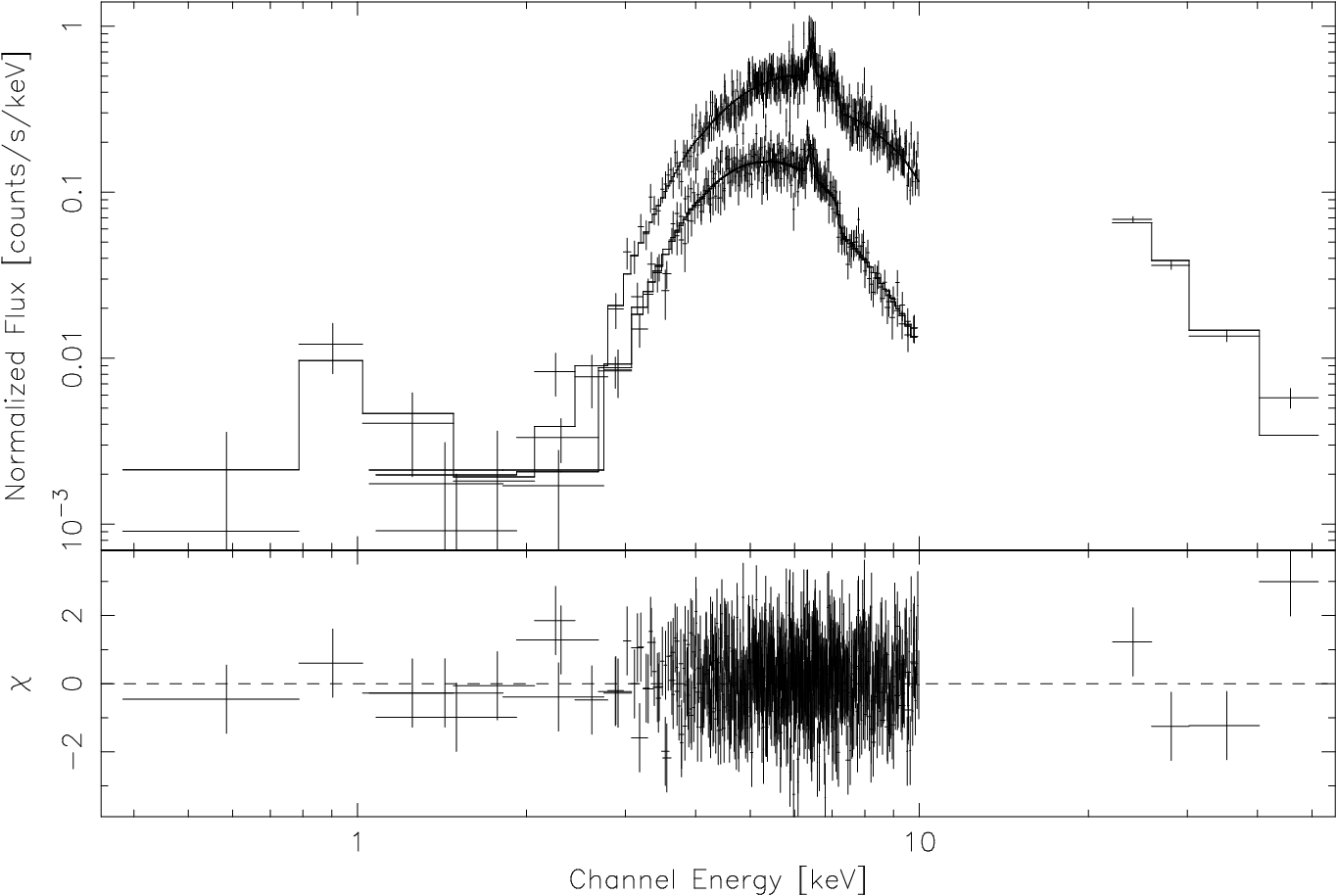}
\caption[]{Photon spectrum of \IGR\ from \PN, \MOSone, \MOStwo, and \ISGRI, fit
with the comptonised continuum model (CE).}
\label{spec_xmm_isgri}
\end{figure}

We fit the combined \EPIC\ and \ISGRI\ spectra (Fig.\,\ref{spec_xmm_isgri}) with a broken power law (BP), a power law with an 
exponential cutoff energy (CP), and a Compton emission (CE) model (\texttt{comptt} within Xspec). The models are modified by a galactic absorption in the direction of the 
source (fixed at $2.2\times10^{22}$cm$^{-2}$ \citep{Dic90}), and an intrinsic photoelectric absorption with free iron abundances 
($\mathrm{Z}_{\mathrm{Fe}}$). Soft excess emission is represented by a blackbody (with a fixed temperature of 
$kT_{b}=0.06$ keV) that is affected by the galactic absorption only. Two narrow Gaussians (with widths fixed at 0) 
describe the iron lines. A constant ($C_{I}$) accounts for the asynchronous observations and for cross-calibration uncertainties.

The BP model ($\chi_{\nu}^{2}/\mathrm{dof}=0.95/675$) has photon indices of $\Gamma_{1}=0.9^{+0.1}_{-0.2}$ and $\Gamma_{2}=4.6^{+0.8}_{-0.3}$. 
However, the BP model does not constrain the break energy ($E_{\mathrm{break}}>$17\,keV) nor 
$C_{I}$ because it requires one more parameter than the other models. The CP model ($\chi_{\nu}^{2}/\mathrm{dof}=1.07/676$) 
gives $\Gamma=0.8\pm0.2$, and a cutoff energy of $E_{\mathrm{cut}}=10\pm1$\,keV, but residuals remain at high energy because of the low 
cutoff temperature. In \citet{Lut05a}, an absorbed cutoff power law fit to the combined \ASCA\ and \ISGRI\ spectra had a comparable cutoff energy of 
11$\pm$1 keV, but the photon index was poorly constrained ($\Gamma=1.3\pm1.0$). 

Parameters from the CE model ($\chi_{\nu}^{2}/\mathrm{dof}=0.95/676$) are listed in Table\,\ref{tab_spec}. The column density ($N_{\mathrm{H}}$) is 
$(25\pm2)\times10^{22}$cm$^{-2}$. The comptonising medium has an electron temperature
($kT_{\mathrm{e}}$) of 4.4$\pm$0.3 keV with an optical depth ($\tau$) of 9$\pm$1. The Fe\,K$_{\alpha}$ line is at 6.41$\pm$0.03
and has an equivalent width of 60$\pm$30 eV when measured with respect to the unabsorbed continuum, while the
energy of the Fe\,K$_{\beta}$ line is 7.1$\pm$0.1 keV with an equivalent width of $<$120 eV. 
The detection of the K$_{\beta}$ line is marginal given the uncertainties and its F-test probability of 3\%. However, the ratio of 
iron intensities ($F_{K\beta}/F_{K\alpha}$) is consistent with the value expected from the photoionisation of iron \citep{Kaa93}.
The absorbed, integrated fluxes (in units of $10^{-11}$ergs\,cm$^{-2}$\,s$^{-1}$) are 4.4 in the 2--10 keV band,
and 5.1 in the 20--60 keV band. The observed soft excess flux between 0.5 and 2 keV is $7\times10^{-15}$ergs\,cm$^{-2}$\,s$^{-1}$. 
Since $C_{\mathit{I}}\sim1$, the \EPIC\ observation occurred during a period in which the source was in an average state.

\subsection{Phase-resolved Spectrum}
\label{spec_phase_res}
The \PN\ spectra were binned according to two states: a pulse state around
the primary peak in the folded light curve (phase 0.43--0.77 in Fig.\,\ref{xmm_var}b), and a plateau state
(phases 0.00--0.07 and 0.87--1.00). Average count rates for the pulse and plateau
states are almost a factor of 2 apart at $\sim$1.2 and $\sim$0.7, respectively. The CE model was applied to
the phase-resolved \EPIC\ spectra constrained with the \ISGRI\ spectrum.
Bins for the plateau gather 20 counts and bins for the pulse collect
40 counts per channel. The soft excess is not prominent in either spectrum
so the blackbody component is omitted. This prevents an evaluation of the influence of the pulse on the
soft excess. Table\,\ref{tab_spec} reports parameters for the CE model fit to the phase-resolved spectra. 
The pulsation affects the normalisations but does not modify the
shape of the spectra nor its parameters, specifically the $N_{\mathrm{H}}$, in any appreciable way.

\begin{figure}[!t]
\psfig{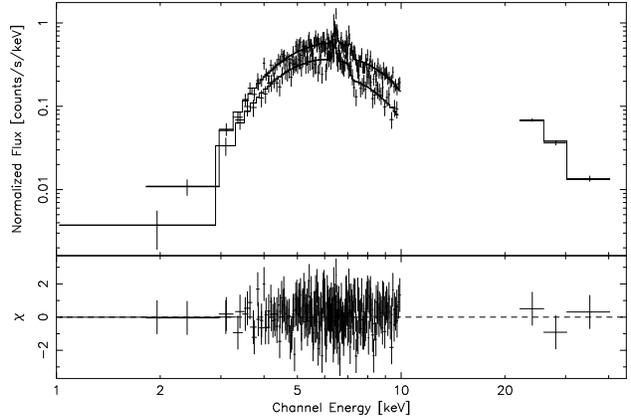}
\caption[]{Phase-resolved spectrum of \IGR\ from \PN\ during the pulse and plateau states.
The \ISGRI\ spectrum is included to constrain the CE model.}
\label{spec_xmm_phase}
\end{figure}

\section{Discussion}
\label{disc}

Observations of \IGR\ by \INT\ and \XMM\ present a source that is highly-obscured and persistent, 
with an intensity varying by a factor larger than 20. The source has a pulsation period of 912.0$\pm$0.1 s. These attributes,
along with its spectral shape, its neutral iron lines, and its lack of radio emission, suggest a HMXB system
consisting of a magnetised neutron star orbiting inside the dense stellar wind of a supergiant companion. In the 
2--10 keV energy band, the unabsorbed flux of \IGR\ is $9.2\times10^{-11}$\,ergs\,cm$^{-2}$s$^{-1}$. Assuming a luminosity of 
$1.2\times10^{36}$\,ergs\,s$^{-1}$, which is typical for accretion-driven X-ray pulsars \citep{Bil97}, the source is located 
approximately 10 kpc away.

\begin{table*}[!t] \centering
        \caption{Parameters from the Compton emission (CE) model fit to the spectra of \EPIC\ and \ISGRI\ combined. 
	Also listed are the values of the CE model (without
	a blackbody) fit to the pulse and plateau states of the phase-resolved \PN\ spectra
	constrained with the \ISGRI\ spectrum. The integrated fluxes are listed as observed ($F_{2-10}$) or unabsorbed
	($F^{\mathrm{UA}}_{2-10}$). Errors denote 90\% confidence.}
        \label{tab_spec}
        \begin{tabular}{lcccl}  \hline
        \noalign{\smallskip}
        \noalign{\smallskip}
	Parameter & CE & Pulse & Plateau & Unit \\
        \noalign{\smallskip}
        \noalign{\smallskip}
	\hline
        \noalign{\smallskip}
        \noalign{\smallskip}
        $C_{I}$ & 0.8$\pm$0.2 & 0.6$\pm$0.2 & 1.3$\pm$0.7 &  \\
	$\tau$ & 9$\pm$1 & 12$^{+6}_{-2}$ & 8$^{+4}_{-2}$ &  \\
        $kT_{\mathrm{e}}$ & 4.4$\pm$0.3 & 4.4$\pm$0.3 & 4.4$\pm$0.4 & keV  \\
        $N_{\mathrm{H}}$ & 25$\pm$2 & 23$^{+2}_{-6}$ & 24$^{+3}_{-7}$ & 10$^{22}$cm$^{-2}$  \\
	$\mathrm{Z}_{\mathrm{Fe}}$ & 1.0$^{+0.4}_{-0.2}$ & 0.9$^{+0.6}_{-0.3}$ & 1.2$^{+0.8}_{-0.6}$ & $\mathrm{Z}_{\odot}$  \\
	$F_{2-10}$ & 4.4 & 5.2 & 3.0 & 10$^{-11}$ergs\,cm$^{-2}$\,s$^{-1}$  \\
	$F^{\mathrm{UA}}_{2-10}$ & 9.2 & 9.8 & 6.7 & 10$^{-11}$ergs\,cm$^{-2}$\,s$^{-1}$  \\
	E$_{K\alpha}$ & 6.41$\pm$0.03 & 6.44$\pm$0.03 & 6.41$\pm$0.04 & keV  \\
	$F_{K\alpha}$ & 1.1$\pm$0.2 & 1.1$\pm$0.4 & 0.8$\pm$0.4 & 10$^{-4}$ph\,cm$^{-2}$\,s$^{-1}$  \\
	E$_{K\beta}$ & 7.1$\pm$0.1 & 7.1$\pm$0.3 & 7.2$\pm$0.3 & keV  \\
	$F_{K\beta}$ & $<$0.62 & $<$1.1 & $<$1.3 & 10$^{-4}$ph\,cm$^{-2}$\,s$^{-1}$  \\
	$\chi_{\nu}^{2}/$d.o.f. & 0.95/676 & 0.92/140 & 0.92/94 &  \\
	\noalign{\smallskip}
        \noalign{\smallskip}
        \hline
        \end{tabular}
\end{table*}
 
The absorbing column density is large whenever the source is detected, whether by \ASCA\ 
($N_{\mathrm{H}}=13^{+9}_{-7}\times10^{22}$cm$^{-2}$), or now with \INT\ and \XMM\ 
($N_{\mathrm{H}}=(25\pm2)\times10^{22}$cm$^{-2}$). The strong absorption below 5 keV is intrinsic since it is 
an order of magnitude larger than the galactic absorption along the 
line of sight ($N_{\mathrm{H}}=2.2\times10^{22}$cm$^{-2}$). The absorption does not vary with the pulsation, which means that it is not 
related to the accretion column, but it may change with the orbital phase. This absorption indicates the presence of optically-thick 
material surrounding the compact object, which is consistent with the detection of iron lines.
 
Emission lines at 6.4 and 7.1 keV trace the quantity of matter in the shell that
envelopes the neutron star. These lines are at the positions that would be expected from the fluorescence of cold neutral 
matter illuminated by continuum X-rays, and their equivalent widths are compatible with a spherical distribution 
of dense gas around the compact object \citep{Mat02}. A K$\alpha$ line at 6.41$\pm$0.03\,keV ($\gtrsim$1.925 $\AA$) corresponds to an ionisation 
of at most Fe XVIII \citep{Hou69}, which does not constrain the distance between the ionising source and the inner surface of the ionised shell. 
Given the errors on the line equivalent widths, we can not determine whether the fluxes of the iron lines are modified by the pulse.

There appears to be a soft excess flux of $7\times10^{-15}$ergs\,cm$^{-2}$\,s$^{-1}$ that is best represented by a blackbody rather than with a 
partial covering or partially-ionised absorber. The scattering of X-rays by the stellar wind is the most 
likely explanation for the soft excess emission \citep{Hab90}. To what extent the pulse affects the soft excess emission is 
still unknown as the soft excess flux is not prominent in the phase-resolved spectrum.

The \ASCA\ Faint Source Catalog \citep{Sug01} listed \IGR\ among the brightest objects in its catalog,
but the uncertainty of the position and the heavy absorption prevented \citet{Sug01} from
associating this source with an optical counterpart. \citet{Mal04}
noted the possible association between \IGR\ and the unidentified \emph{EGRET} source 3EG\,J1639$-$4702 \citep{Har99}. 
Non-thermal radio emission was recently detected within the \ASCA\ error box \citep{Com04} which suggests 
a dust-enshrouded microquasar and could help justify the \emph{IBIS} and \emph{EGRET} association. 
However, the \XMM\ position (R.A. (J2000)$=16^{\mathrm{h}}39^{\mathrm{m}}05.4^{\mathrm{s}}$, Dec.$=-46^{\circ}42'12''$, 4$''$ 
uncertainty) is incompatible with the position of the radio source MOST\,J1639.0$-$4642, or any known radio source 
from the Vizier database. The pulsation and the lack of radio emission invalidate the microquasar interpretation for the HMXB. 
Also, the chance association of the \emph{EGRET} source with 
\IGR\ is strong given the high density of sources in the region. 

A single infrared candidate, 2MASS\,J16390535$-$4642137, lies within the \XMM\ error circle of 
IGR\,J16393$-$4643. The high-energy spectral and timing characteristics of \IGR\ are reminiscent of other heavily-absorbed, 
wind-accreting X-ray pulsars with OB supergiant companions \citep{Wal04}. A Be stellar companion is unlikely 
given that such systems are usually transient. Infrared observations should 
confirm the supergiant nature of the companion. If this is the case, it will constitute another argument rejecting the 
\emph{EGRET} association \citep{Ore05}.

Recently, combined \INT\ and \XMM\ observations of the Norma Arm have revealed more pulsating X-ray binaries than 
were previously known. These objects have joined the ranks of what might be a new class of heavily-absorbed HMXBs 
that were previously undetected below 5 keV. The increasing sample size these objects represent should enable meaningful 
statistical studies to be performed. Understanding the nature of sources such as \IGR\ could shed light on the structure of 
stellar winds, provide constraints on the masses of neutron stars, and help elucidate the evolution of binaries.

\begin{acknowledgements}
The authors thank the anonymous referee for their prompt report which improved the paper. 
A. Bodaghee acknowledges J. Wendt, J. Rodriguez, A. Paizis, D. Willis, N. Produit and S. Paltani for their input and discussions. 
This research has made use of the SIMBAD database, operated at CDS, Strasbourg, France. This publication uses 
observations obtained with the ESA science missions \INT\ and \XMM. The \INT\ and \XMM\ instruments and 
data centres were directly funded by ESA member states and the USA (NASA).
\end{acknowledgements}

\bibliographystyle{aa}
\bibliography{bodaghee.bib}

\end{document}